\documentclass[a4paper,10pt]{article}
\def\int {\intop \limits}

\usepackage{mathptmx}
\usepackage{wrapfig}
\usepackage{graphicx}
\usepackage[superscript]{cite}
\usepackage[left=19.25mm,right=19.25mm,top=25.4mm,bottom=49.4mm]{geometry}

\begin{document}

\begin{center}
\textbf{\fontsize{16}{20pt}\selectfont Coherent and incoherent
processes and the LPM}\\
\textbf{\fontsize{16}{20pt}\selectfont effect in oriented single
crystals at high-energy}\\[16pt]
{\fontsize{12}{14pt}\selectfont V.~N.~Baier\footnote{V.N.Baier@inp.nsk.su}, V.~M.~Katkov}\\
{\fontsize{12}{14pt}\selectfont Budker Institute of Nuclear Physics,}\\
{\fontsize{12}{14pt}\selectfont Novosibirsk, 630090, Russia}\\[24pt]
\textbf{\fontsize{12}{14pt}\selectfont ABSTRACT}
\end{center}

The process of radiation from high-energy electron and
electron-positron pair production by a photon in oriented single
crystal is considered using the method which permits inseparable
consideration of both coherent and incoherent mechanisms of photon
emission from an electron and of pair creation by a photon and
includes the action of field of axis (or plane) as well as the
multiple scattering of radiating electron or particles of the
created pair (the Landau-Pomeranchuk-Migdal (LPM) effect). The total
intensity of radiation and total probability pair creation are
calculated. The theory, where the energy loss of projectile has to
be taken into account, and found probabilities of pair creation
agree quite satisfactory with available CERN data. From obtained
results it follows that multiple scattering appears only for
relatively low energy of radiating electron or a photon, while at
higher energies the field action excludes the LPM effect.

\vspace{2mm}
\begin{center}
\textbf{\fontsize{11}{13pt}\selectfont 1.~ INTRODUCTION}
\end{center}
\vspace{2mm}

The mechanisms of basic electromagnetic process in oriented single
crystal (radiation and pair production by a photon)  differ
substantially from the mechanisms of independent photon emission
from electron or pair  photoproduction at separate centers acting in
an amorphous medium (the Bethe-Heitler mechanisms). In crystal the
coherent interaction of an electron (or a photon) with many centers
occurs. Under some generic assumptions the general theory of the
coherent radiation mechanism was developed in \cite{BKS0} (of the
coherent pair creation mechanism in \cite{BKS1}). Recently authors
developed the new approach to analysis of pair creation by a photon
\cite{BK0} and radiation \cite{BK1} in oriented crystals . This
approach not only permits indivisible consideration of both the
coherent and incoherent mechanisms of radiation (or pair creation by
a photon) but also gives insight on the Landau-Pomeranchuk-Migdal
(LPM) effect (influence of multiple scattering) on the considered
processes.

The properties of process are connected directly with details of
motion of emitting particle (or particles of created pair). The
momentum transfer from a particle to a crystal we present in a form
${\bf q}=<{\bf q}>+{\bf q}_s$, where $<{\bf q}>$ is the mean value
of momentum transfer calculated with averaging over thermal(zero)
vibrations of atoms in a crystal. The motion of particle in an
averaged potential of crystal, which corresponds to the momentum
transfer $<{\bf q}>$, determines the coherent mechanism of process.
The term ${\bf q}_s$ is attributed to the random collisions of
particle which define the incoherent radiation (pair creation). Such
random collisions we will call "scattering" since $<{\bf q}_s>=0$.
If the formation length of the process is large with respect to
distances between atoms forming the axis, the additional averaging
over the atom position should be performed.

If the electron (or photon) angle of incidence $\vartheta_0$ (the
angle between the electron momentum {\bf p} (or photon momentum {\bf
k}) and the axis (or plane)) is small $\vartheta_0 \ll V_0/m$, where
$V_0$ is the characteristic scale of the potential, the field $E$ of
the axis (or plane) can be considered constant over the process
formation length and the constant-field approximation is valid. In
this case the behavior of radiation probability is determined by the
parameter $\chi$ and the pair production probability is determined
by the parameter $\kappa$:
\begin{equation}
\chi=\frac{\varepsilon}{m}\frac{E}{E_0},
\quad\kappa=\frac{\omega}{m}\frac{E}{E_0},
\label{1}
\end{equation}
where $\varepsilon(\omega)$ is the electron (photon) energy, $m$ is
the electron mass, $E_0=m^2/e=1.32 \cdot 10^{16}$ V/cm is the
critical field, the system $\hbar=c=1$ is used.

The very important feature of coherent radiation (or coherent pair
creation) mechanism is the strong enhancement of its probability at
high energies (from factor $\sim 10$ for main axes in crystals of
heavy elements like tungsten to factor $\sim 160$ for diamond)
comparing with the Bethe-Heitler mechanism which takes place in an
amorphous medium. If $\vartheta_0 \gg V_0/m$ the theory passes over
to the coherent bremsstrahlung theory or the coherent pair
production theory (see \cite{D, T, BKS}). Side by side with coherent
mechanism the incoherent mechanism of radiation is acting. In
oriented crystal this mechanism changes also with respect to an
amorphous medium \cite{BKS2}. The details of theory and description
of experimental study of radiation and pair creation which confirms
the mentioned enhancement can be found in \cite{BKS}. The study of
radiation and pair creation in oriented crystals is continuing and
new experiments are performed recently \cite{KMU, BKi, KKM, B}.

At high energies the multiple scattering of radiating electron or
particles of created pair (the LPM effect) suppresses radiation (or
pair creation) probability when $\varepsilon \geq \varepsilon_e$ (or
$\omega \geq \omega_e$). In an amorphous medium (or in crystal in
the case of random orientation) the characteristic electron energy
starting from which the LPM effect becomes essential is
$\varepsilon_e \sim 2.5$~TeV  for heavy elements \cite{BK10} and
this value is inversely proportional to the density. In the vicinity
of crystalline axis (just this region gives the crucial contribution
to the Bethe-Heitler mechanism) the local density of atoms is much
higher than average one and for heavy elements and at low
temperature the gain could attain factor $\sim 10^3$. So in this
situation the characteristic electron energy can be $\varepsilon_0
\sim 2.5$~GeV and this energy is significantly larger than
"threshold" energy $\varepsilon_t$ starting from which the
probability of coherent radiation exceeds the incoherent one. For
pair photoproduction the characteristic photon energies are 4 times
larger: $\omega_e=4\varepsilon_e \sim 10$~TeV for heavy elements in
an amorphous medium and in crystal $\omega_0 = \omega_e/\xi(0) \sim
10$~GeV. The last energy is of the order of the threshold energy
$\omega_t$ for which the probability of pair creation in the axis
field becomes equal to the Bethe-Maximon probability, see Sec.12.2
and Table 12.1 in \cite{BKS}. It should be noted that the main
contribution into the multiple scattering gives the small distance
from axis where the field of crystalline axis attains the maximal
value. For the same reason the LPM effect in oriented crystals
originates in the presence of crystal field and nonseparable from
it. This means that in problem under consideration we have both the
dense matter with strong multiple scattering and high field of
crystalline axis.

Below we consider case $\vartheta_0 \ll V_0/m$. Than the distance of
an electron from axis $\mbox{\boldmath$\varrho$}$ as well as the
transverse field of the axis can be considered as constant over the
formation length. For an axial orientation of crystal the ratio of
the atom density $n(\varrho)$ in the vicinity of an axis to the mean
atom density $n_a$ is
\begin{equation}
\frac{n(x)}{n_a}=\xi(x)=\frac{x_0}{\eta_1}e^{-x/\eta_1},\quad
\varepsilon_0=\frac{\varepsilon_e}{\xi(0)}, \label{2}
\end{equation}
where
\begin{equation}
x_0=\frac{1}{\pi d n_a a_s^2}, \quad  \eta_1=\frac{2
u_1^2}{a_s^2},\quad x=\frac{\varrho^2}{a_s^2}, \label{3}
\end{equation}
Here $\varrho$ is the distance from axis, $u_1$ is the amplitude of
thermal vibration, $d$ is the mean distance between atoms forming
the axis, $a_s$ is the effective screening radius of the axis
potential (see Eq.(9.13) in \cite{BKS})
\begin{equation}
U(x)=V_0\left[\ln\left(1+\frac{1}{x+\eta} \right)-
\ln\left(1+\frac{1}{x_0+\eta} \right) \right]. \label{4}
\end{equation}
The local value of parameters $\chi(x)$  ($\kappa(x)$), see
Eq.(\ref{1}), which determines the radiation (pair creation)
probability in the field Eq.(\ref{4}) is
\begin{equation}
\chi(x)=-\frac{dU(\varrho)}{d\varrho}\frac{\varepsilon}{m^3}=\chi_s
f_a,\quad f_a=\frac{2\sqrt{x}}{(x+\eta)(x+\eta+1)},\quad
\chi_s=\frac{V_0 \varepsilon}{m^3a_s}\equiv
\frac{\varepsilon}{\varepsilon_s},\quad \kappa(x)=\kappa_sf_a,\quad
\kappa_s=\frac{V_0 \omega}{m^3a_s}\equiv \frac{\omega}{\omega_s}.
 \label{5}
\end{equation}

The parameters of the axial potential for the ordinarily used
crystals are given in Table 9.1 in \cite{BKS}. The particular
calculation below will be done for tungsten and germanium crystals
studied in \cite{KMU, KKM}. The relevant parameters are given in
Table 1.

\begin{table}[h]
\begin{center}
{\sc Table 1}~ {Parameters of radiation (pair creation) process in
the tungsten (the axis $<111>$)\\ and germanium (the axis $<110>$)
crystals for different temperatures T,  the energies~ $\varepsilon$
and $\omega$~ are in GeV}
\end{center}
\begin{center}
\begin{tabular}{*{12}{|c}|}
\hline Crystal &T(K)&$V_0$(eV)&$x_0$&$\eta_1$&$\eta$&
$\varepsilon_0$&$\varepsilon_t$&$\varepsilon_s(\omega_s)$&$\varepsilon_m(\omega_m)$&$\omega_0$&$h$ \\
\hline W&293&413&39.7&0.108&0.115&7.43&0.76&34.8&14.35&29.7&0.348\\
\hline W&100&355&35.7&0.0401&0.0313&3.06&0.35&43.1&8.10&12.25&0.612\\
\hline Ge&100&114.5&19.8&0.064&0.0633&59&0.85&179&51&236&0.459\\
\hline
\end{tabular}
\end{center}
\end{table}

\newpage

\begin{center}
\textbf{\fontsize{11}{13pt}\selectfont 2.~ PROCESSES IN LIMITING
CASES}
\end{center}
\textbf{\fontsize{10}{12pt}\selectfont 2.1.~Radiation}

It is useful to compare the characteristic energy $\varepsilon_0$
(or $\omega_0$ for pair creation) with "threshold" energy
$\varepsilon_t$ (or $\omega_t$ for pair creation) for which the
radiation intensity (pair creation probability) in the axis field
becomes equal to the Bethe-Maximon one. Since the maximal value of
parameter $\chi(x)$ :
\begin{equation}
\chi_m=\chi(x_m),\quad \kappa_m=\kappa(x_m),\quad
x_m=\frac{1}{6}(\sqrt{1+16\eta(1+\eta)}-1-2\eta),\quad
\chi_m=\frac{\varepsilon}{\varepsilon_m},\quad
\kappa_m=\frac{\omega}{\omega_m} \label{6}
\end{equation}
is small for such electron energy $(\varepsilon_t \ll
\varepsilon_m)$, one can use the decomposition of radiation
intensity over powers of $\chi$ (see Eq.(4.52) in \cite{BKS}) and
carry out averaging over $x$. Retaining three terms of decomposition
we get
\begin{eqnarray}
&&I^F = \frac{8\alpha m^2
\chi_s^2}{3x_0}\left(a_0(\eta)-a_1(\eta)\chi_s +
a_2(\eta)\chi_s^2+\ldots \right), \
\nonumber \\
&& a_0(\eta)=(1+2\eta)\ln \frac{1+\eta}{\eta}-2, \quad
 a_1(\eta)=\frac{165 \sqrt{3}
\pi}{64}\left[\frac{1}{\sqrt{\eta}}-\frac{1}{\sqrt{1+\eta}}-
4\left(\sqrt{1+\eta} - \sqrt{\eta} \right)^3\right],
\nonumber \\
&& a_2(\eta)=64\left[(1+2\eta)\left(\frac{1}{\eta(1+
\eta)}+30\right)- 12(1+5\eta(1+\eta))\ln \frac{1+\eta}{\eta}\right].
\label{7}
\end{eqnarray}

The intensity of incoherent radiation in low energy region
$\varepsilon \leq \varepsilon_t \ll \varepsilon_m$ is (see
Eq.(21.16) in \cite{BKS} and Eq.(\ref{a.16}) below)
\begin{equation}
I^{inc}=\frac{\alpha
m^2}{4\pi}\frac{\varepsilon}{\varepsilon_e}g_{0r}\left[1+34.4
\left(\overline{\chi^2\ln \chi}+
2.54\overline{\chi^2}\right)\right], \quad
g_{0r}=1+\frac{1}{L_0}\left[\frac{1}{18}-h\left(\frac{u_1^2}{a^2}\right)\right],
\quad \overline{f} = \int_0^{\infty} f(x)
e^{-\frac{x}{\eta_1}}\frac{dx}{\eta_1}, \label{8}
\end{equation}
where
\begin{eqnarray}
&& \varepsilon_e=\frac{m}{16\pi Z^2\alpha^2\lambda_c^3n_aL_0},\quad
L_0=\ln(ma)+ \frac{1}{2}-f(Z\alpha), \quad
h(z)=-\frac{1}{2}\left[1+(1+z)e^{z}{\rm Ei}(-z) \right],\quad
a=\frac{111Z^{-1/3}}{m},
\nonumber \\
&& f(\xi)={\rm Re} \left[\psi(1+i\xi)-\psi(1)
\right]=\sum_{n=1}^{\infty} \frac{\xi^2}{n(n^2+\xi^2)}, \label{9}
\end{eqnarray}
here $\psi(z)$ is the logarithmic derivative of the gamma function,
Ei($z$) is the integral exponential function, $f(\xi)$ is the
Coulomb correction. For $\chi=0$ this intensity differs from the
Bethe-Maximon intensity only by the term $h(u_1^2/a^2)$ which
reflects the nonhomogeneity of atom distribution in crystal. For
$u_1\ll a$ one has $h(u_1^2/a^2) \simeq
-(1+C)/2+\ln(a/u_1),~C=0.577..$ and so this term characterizes the
new value of upper boundary of impact parameters $u_1$ contributing
to the value $<{\bf q}_s^2>$ instead of screening radius $a$ in an
amorphous medium.

Conserving in Eq.(\ref{7}) only the main (the first) term of
decomposition, which corresponds to the classical radiation
intensity, neglecting the corrections in Eq.(\ref{8})
($g_0=1,~\chi=0$), using the estimate $V_0 \simeq Z\alpha/d$ and
Eqs.(\ref{3}), (\ref{5}), we get
\begin{equation}
\varepsilon_t \simeq \frac{3 L_0 d m^2}{2\pi a_0(\eta)} =
63\frac{L_0 d}{a_0(\eta)}{\rm MeV}, \label{10}
\end{equation}
where the distance $d$ is taken in units $10^{-8}$ cm. Values of
$\varepsilon_t$ found using this estimate for tungsten, axis
$<111>$, $d$=2.74 $\cdot 10^{-8}$ cm are consistent with points of
intersection of coherent and incoherent intensities in Fig.1 (see
Table 1). For some usable crystals (axis $<111>$, room temperature)
one has from Eq.(\ref{10})
\begin{equation}
\varepsilon_t({\rm C_{(d)}}) \simeq 0.47~ {\rm GeV},\quad
\varepsilon_t({\rm Si}) \simeq 2.0~ {\rm GeV},\quad
\varepsilon_t({\rm Ge}) \simeq 1.7~ {\rm GeV}, \label{11}
\end{equation}
so this values of $\varepsilon_t$ are somewhat larger than in
tungsten except the diamond very specific crystal where value of
$\varepsilon_t$ is close to tungsten one.

For large values of the parameter $\chi_m~(\varepsilon \gg
\varepsilon_m)$ the incoherent radiation intensity is suppressed due
to the action of the axis field. In this case the local intensity of
radiation can by written as (see Eq.(7.129) in \cite{BKS})
\begin{equation}
I^{inc}=\frac{29
\Gamma(1/3)}{3^{1/6}2430}\frac{\varepsilon}{\varepsilon_e}\frac{\alpha
m^2}{\chi^{2/3}(x)}\left[g_{0r}+\frac{1}{L_0}\left( 0.727+\frac{\ln
\chi(x)}{3} \right)\right]. \label{12}
\end{equation}
Here we have taken into account that
\begin{equation}
\ln \frac{1}{\gamma \vartheta_1}=\ln(ma) \rightarrow
\ln(ma)-h\left(\frac{u_1^2}{a^2}\right)-f(Z\alpha)=L_0
-h\left(\frac{u_1^2}{a^2}\right)-\frac{1}{2}. \label{12a}
\end{equation}
 Averaging the function $(\chi(x))^{-2/3}$ and
$\ln\chi(x)(\chi(x))^{-2/3}$ over $x$ according with Eq.(\ref{8})
one can find the effective value of upper boundary of the transverse
momentum transfer ($\propto m\chi^{1/3}_m$ instead of $m$) which
contributes to the value $<{\bf q}_s^2>$. Using the obtained results
we determine the effective logarithm $L$ by means of interpolation
procedure
\begin{equation}
L=L_0g_r,\quad g_r=g_{0r}+\frac{1}{6 L_0}\ln
\left(1+70\chi_m^2\right). \label{13}
\end{equation}
Let us introduce the local characteristic energy of electron (see
Eq.(\ref{2}))
\begin{equation}
\varepsilon_c(x)=
\frac{\varepsilon_e(n_a)}{\xi(x)g_r}=\frac{\varepsilon_0}{g_r}e^{x/\eta_1},
\label{14}
\end{equation}
In this notations the contribution of multiple scattering into the
local intensity for small values of $\chi_m$ and
$\varepsilon/\varepsilon_0$ has a form (see Eq.(15) in \cite{BK3}
and Eq.(\ref{d.8}) below)
\begin{equation}
I^{LPM}(x)=-\frac{\alpha m^2}{4\pi}
\frac{\varepsilon}{\varepsilon_c(x)}\left[\frac{4\pi\varepsilon}{15
\varepsilon_c(x)}\left(1+\frac{171\sqrt{3}}{16}\chi(x)\right)+\frac{64
\varepsilon^2}{21 \varepsilon_c^2(x)}\left(\ln
\frac{\varepsilon}{\varepsilon_c(x)}+2.04\right)\right]. \label{15}
\end{equation}
Integrating this expression over $x$ with the weight $1/x_0$ we get
\begin{equation}
I^{LPM}=\frac{\alpha m^2}{4\pi}
\frac{\varepsilon}{\varepsilon_e}g_r\left[-\frac{2\pi\varepsilon
g_r}{15
\varepsilon_0}\left(1+37\mu\right)+\frac{64}{63}\frac{\varepsilon^2
g_r^2}{ \varepsilon_0^2}\left(\ln \frac{\varepsilon_0}{\varepsilon
g_r}-1.71\right)\right],\quad
\mu=\int_0^{\infty}e^{-2x/\eta_1}\chi(x)\frac{dx}{\eta_1}.
\label{16}
\end{equation}
It should be noted that found Eq.(\ref{16}) has a good accuracy only
for energy much smaller (at least on one order of magnitude) than
$\varepsilon_0$ (see discussion after Eq.(15) in \cite{BK3}).

\vspace{2mm}
\textbf{\fontsize{10}{12pt}\selectfont 2.2.~Pair
creation}

For small value of the parameter $\kappa$ the probability of
coherent pair creation is (see Eq.(12.11) in \cite{BKS})
\begin{equation}
W^F=\frac{9}{32}\sqrt{\frac{\pi}{2}}\frac{\alpha m^2}{\omega x_0}
\frac{\kappa_m^2}{\sqrt{-\kappa''_m}}\exp(-8/3\kappa_m), \label{5a}
\end{equation}
where $\kappa_m$ (which defines the value of $\omega_t$) is given in
Eq.(\ref{6}), $\kappa''_m=\kappa''(x_m)$. We find that $\omega_t
\sim \omega_m \sim \omega_0$ for main axes of crystals of heavy
elements. So at $\omega \sim \omega_t$ all the discussed effects are
simultaneously essential in these crystals. In crystals of elements
with intermediate $Z$ (Ge, Si, diamond) the ratio $\omega_t/\omega_m
\sim 1$ but $\omega_m/\omega_0 \ll 1$. So, the LPM effect for such
crystals is significantly weaker.

At $\omega \ll \omega_t$ the  incoherent mechanism of pair creation
dominates. It integral cross section in oriented crystal has the
form (see Eq.(26.30) in \cite{BKS})
\begin{equation}
\sigma_p=\frac{28Z^2\alpha^3}{9m^2}\left[L_0-\frac{1}{42}
-h\left(\frac{u_1^2}{a^2} \right)\right], \label{7p}
\end{equation}
where notations see in Eq.(\ref{9}).

The influence of axis field on the incoherent pair creation process
begins when $\omega$ becomes close to $\omega_m$. For small values
of the parameter $\kappa_m$ the correction to the cross section
Eq.(\ref{6}) is (see Eq.(7.137) in \cite{BKS})
\begin{equation}
 \Delta \sigma_{p}=\frac{176}{175}\frac{Z^2\alpha^3}{m^2}
\overline{\kappa^2} \left(L_u-\frac{1789}{1980}\right),\quad
\overline{\kappa^2}=\int_{0}^{\infty}\frac{dx}{\eta_1}
e^{-x/\eta_1}\kappa^2(x),\quad
L_u=L_0-h\left(\frac{u_1^2}{a^2}\right).
 \label{8p}
\end{equation}

The coherent and incoherent contribution to pair creation can
separated also for $\kappa_m \gg 1~(\omega \gg \omega_m)$. In this
case one can use the perturbation theory in calculation of the
probability of incoherent process and neglect the LPM effect because
of domination of the coherent contribution and additional
suppression (by the axis field) the incoherent process. In this case
the local cross section of pair creation has the form (see
Eq.(7.138) in \cite{BKS})
\begin{equation}
\sigma_p(x)=\frac{8Z^2\alpha^3 \Gamma^3(1/3)}{25
m^2(3\kappa(x))^{2/3}\Gamma(2/3)}
\left(L_u+0.4416+\frac{1}{3}\ln\kappa(x)\right). \label{9p}
\end{equation}
Averaging the function $(\kappa(x))^{-2/3}$ and
$\ln\kappa(x)(\kappa(x))^{-2/3}$ over $x$ according with
Eq.(\ref{8p}) one can find the effective value of upper boundary of
the transverse momentum transfer ($\propto m\kappa^{1/3}_m$ instead
of $m$) which contributes to the value $<{\bf q}_s^2>$. Using the
obtained results we determine the effective logarithm $L$ by means
of interpolation procedure
\begin{equation}
L=L_0g,\quad g=1+\frac{1}{L_0}\left[-\frac{1}{42}-h
\left(\frac{u_1^2}{a^2}\right) +\frac{1}{3}\ln\left
(\frac{6-3\kappa^2_m+3\kappa^3_m}{6+\kappa^2_m}\right) \right].
 \label{10p}
\end{equation}

Let us introduce the local characteristic energy of photon
\begin{equation}
\omega_c(x)=\frac{m}{4\pi Z^2\alpha^2\lambda_c^3n(x)L}=
\frac{\omega_e(n_a)}{\xi(x)g}=\frac{\omega_0}{g}e^{x/\eta_1},
\label{11p}
\end{equation}
where $\lambda_c=1/m$. In this notations the local probability for
small values of $\kappa_m$ and $\omega/\omega_0$ has a form (see
Eq.(7.137) in \cite{BKS} and Eq.(2.23) in \cite{BK10})
\begin{equation}
W(x)=\frac{7}{9\pi}\frac{\alpha m^2}{\omega_c(x)}
\left[1+\frac{396}{1225}\kappa^2(x)-\frac{3312}{2401}
\frac{\omega^2}{\omega^2_c(x)} \right], \label{12p}
\end{equation}
where the term with $\kappa^2(x)$ arises due to the field action and
the term with $\omega^2/\omega^2_c(x)$ reflects influence of
multiple scattering (the LPM effect). Averaging this expression over
$x$ we have
\begin{eqnarray}
&& \int_0^{\infty}\frac{dx}{x_0}\frac{1}{\omega_c(x)}=
\frac{g}{\omega_0}\frac{\eta_1}{x_0}=\frac{g}{\omega_e(n_a)},\quad
\int_0^{\infty}\frac{dx}{x_0}\frac{1}{\omega^3_c(x)}=
\frac{g}{\omega_e(n_a)}\frac{g^2}{3\omega_0^2},
\nonumber \\
&& \int_0^{\infty}\frac{dx}{x_0}\frac{\kappa^2(x)}{\omega_c(x)}=
\frac{g}{\omega_e(n_a)}\overline{\kappa^2}, \quad
 W \equiv \overline{W(x)}=W_0 g
\left[1+\frac{396}{1225}\overline{\kappa^2}
-\frac{1104}{2401}\left(\frac{\omega g}{\omega_0} \right)^2 \right],
 \label{13p}
\end{eqnarray}
where $W_0$ is
\begin{equation}
W_0=\frac{7}{9}\frac{\alpha m^2}{\pi \omega_e(n_a)}=
\frac{28}{9}\frac{Z^2\alpha^3}{m^2}n_a L_0. \label{14p}
\end{equation}

\begin{center}
\textbf{\fontsize{11}{13pt}\selectfont 3.~ GENERAL THEORY}
\end{center}
\textbf{\fontsize{10}{12pt}\selectfont 3.1.~Radiation}

The spectral probability of radiation under the simultaneous action
of multiple scattering and an external constant field was derived in
\cite{BKS} (see Eqs.(7.89) and (7.90)). Multiplying the expression
by $\omega$ and integrating over $\omega$ one obtains the total
intensity of radiation $I$. For further analysis and numerical
calculation it is convenient to carry out some transformations
\begin{enumerate}
\item Changing of variables:~$\nu \rightarrow a\nu/2,~ \tau \rightarrow
2t/a,~(\nu\tau \rightarrow \nu t)$.
\item
Turn the contour of integration over $t$ at the angle $-\pi/4$.
\end{enumerate}
One finds after substitution $t \rightarrow \sqrt{2}t$
\begin{eqnarray}
&& I(\varepsilon)=\frac{\alpha m^2}{2\pi}\int_0^1 \frac{y dy}{1-y}
\int_0^{x_0}\frac{dx}{x_0}G_r(x, y),\quad G_r(x, y)=\int_0^{\infty}
F_r(x, y, t)dt -r_3\frac{\pi}{4}, \quad
y=\frac{\omega}{\varepsilon},
\nonumber \\
&& F_r(x, y, t)={\rm Im}\left\lbrace
e^{\varphi_1(t)}\left[r_2\nu_0^2
(1+ib_r)\varphi_2(t)+r_3\varphi_3(t) \right] \right\rbrace,\quad
b_r=\frac{4\chi^2(x)}{u^2\nu_0^2}, \quad u=\frac{y}{1-y},
\nonumber \\
&&  \varphi_1(t)=(i-1)t+b_r(1+i)(\varphi_2(t)-t),\quad
\varphi_2(t)=\frac{\sqrt{2}}{\nu_0}\tanh\frac{\nu_0t}{\sqrt{2}},\quad
\varphi_3(t)=\frac{\sqrt{2}\nu_0}{\sinh(\sqrt{2}\nu_0t)}, \label{17}
\end{eqnarray}
where
\begin{equation}
r_2=1+(1-y)^2,\quad r_3=2(1-y),\quad \nu_0^2=\frac{1-y}{y}
\frac{\varepsilon}{\varepsilon_c(x)}, \label{18}
\end{equation}
$\omega$ is the photon energy, the function $\varepsilon_c(x)$ is
defined in Eq.(\ref{14}) and $\chi(x)$ is defined in Eq.(\ref{5}).

\begin{wrapfigure}{l}{0.42\textwidth}
\includegraphics[width=0.42\textwidth]{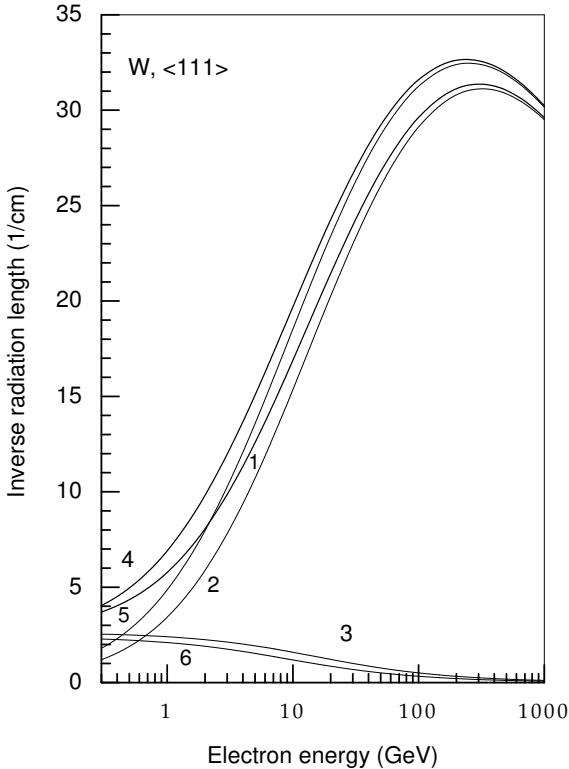}
\caption{The inverse radiation length in tungsten, axis $<111>$ at
different temperatures T vs the electron initial energy. Curves 1
and 4 are the total effect:
$L^{cr}(\varepsilon)^{-1}=I(\varepsilon)/\varepsilon$ Eq.(\ref{17})
for T=293 K and T=100 K correspondingly, the curves 2 and 5 give the
coherent contribution $I^F(\varepsilon)/\varepsilon$ Eq.(\ref{19}),
the curves 3 and 6 give the incoherent contribution
$I^{inc}(\varepsilon)/\varepsilon$ Eq.(\ref{21}) at corresponding
temperatures T.}
\end{wrapfigure}

In order to single out the influence of the multiple scattering (the
LPM effect) on the process under consideration, we should consider
both the coherent and incoherent contributions. The probability of
coherent radiation is the first term ($\nu_0^2=0$) of the
decomposition of Eq.(\ref{17}) over $\nu_0^2$.  The coherent
intensity of radiation is (compare with Eq.(17.7) in \cite{BKS})
\begin{equation}
I^{F}(\varepsilon)=\int_0^{x_0}I(\chi)\frac{dx}{x_0}. \label{19}
\end{equation}
Here $I(\chi)$ is the radiation intensity in constant field
(magnetic bremsstrahlung limit, see Eqs. (4.50), (4.51) in
\cite{BKS}). It is convenient to use the following representation
for $I(\chi)$
\begin{eqnarray}
&&I(\chi)=i\frac{\alpha m^2}{2\pi}
\int_{\lambda-i\infty}^{\lambda+i\infty}
\left(\frac{\chi^2}{3}\right)^s \Gamma\left(1-s\right)
\Gamma\left(3s-1\right)
\nonumber \\
&&\times (2s-1) (s^2-s+2)\frac{ds}{\cos\pi s},\quad
\frac{1}{3}<\lambda<1. \label{20}
\end{eqnarray}

The intensity of incoherent radiation is the second term ($\propto
\nu_0^2$) of the mentioned decomposition. The expression for the
intensity of incoherent radiation follows from Eq.(21.21) in
\cite{BKS}):
\begin{equation}
I^{inc}(\varepsilon)=\frac{\alpha m^2}{60\pi}
\frac{\varepsilon}{\varepsilon_0}g\int_0^{x_0}e^{-x/\eta_1}J(\chi)\frac{dx}{x_0},
\label{21}
\end{equation}
here $J(\chi)$ is the integral over photon energy $\omega$:
\begin{eqnarray}
&&J(\chi)=\int_0^{1}\left[y^2(f_1(z)+f_2(z))+2(1-y)f_2(z)\right]dy,
\nonumber \\
&&z=\left(\frac{y}{\chi(1-y)}\right)^{2/3}, \label{a.1}
\end{eqnarray}
where $y=\omega/\varepsilon$, the functions $f_1(z)$ and $f_2(z)$
are defined in the just mentioned equation in \cite{BKS}:
\begin{eqnarray}
&& f_1(z)=z^4\Upsilon(z)-3z^2\Upsilon'(z)-z^3,
\nonumber \\
&& f_2(z)=(z^4+3z)\Upsilon(z)-5z^2\Upsilon'(z)-z^3, \label{a.11a}
\end{eqnarray}
here $\Upsilon(z)$ is the Hardy function:
\begin{equation}
\Upsilon(z)=\int_0^{\infty}\sin\left(z\tau+\frac{\tau^3}{3}\right)d\tau.
\label{a.12a}
\end{equation}
We used the following relations between the function $\Upsilon(z)$
and its derivatives:
\begin{eqnarray}
&& \Upsilon^{(n)}=\frac{d^n}{dz^n}{\rm Im}\int_0^{\infty}
\exp\left(i\left(z\tau+\frac{\tau^3}{3}\right)\right)d\tau ={\rm
Im}\int_0^{\infty}(i\tau)^n
\exp\left(i\left(z\tau+\frac{\tau^3}{3}\right)\right)d\tau,
\nonumber \\
&& z\Upsilon(z)=\Upsilon''(z)+1,\quad
\Upsilon^{(n+3)}(z)=(n+1)\Upsilon^{(n)}+z\Upsilon^{(n+1)}.
\label{a.6}
\end{eqnarray}
Integrating Eq.(\ref{a.1}) by parts one can represent the integral
$J(\chi)$ in the form
\begin{eqnarray}
&& J(\chi)
=\frac{\chi^3}{6}\frac{d^2}{d\chi^2}(J_1(\chi)+J_2(\chi))+\frac{d}{d\chi}(\chi^2J_2(\chi)),
\quad
 J_{1,2}(\chi)=\frac{f_{1,2}(\infty)}{\chi}+i_{1,2}(\chi),
\nonumber \\
&& i_{1,2}(\chi)=\chi\int_0^{\infty}f_{1,2}'(z)\frac{z^3dz}{1+\chi
z^{3/2}}. \quad
 f_{1}'(z)=z^2\Upsilon^{(5)}(z)-3z\Upsilon^{(4)},\quad
f_{2}'(z)=z^2\Upsilon^{(5)}(z)-5z\Upsilon^{(4)}+3\Upsilon^{(3)}.
\label{a.9}
\end{eqnarray}
Since the integrals in  Eq.(\ref{a.9}) for the separate terms of
functions $f_{1.2}'(z)$ in form Eq.(\ref{a.11a}) diverges, we
transformed it to the form Eq.(\ref{a.9}). We used also the
important formula
\begin{equation}
\int_0^{\infty} z^{3/2}f_{1,2}'(z)dz=0, \label{a.7}
\end{equation}
which follows from the equation $\int_0^{\infty}
\Upsilon'(z)\frac{dz}{\sqrt{z}}=0$ having applied integration by
parts for separate terms of functions $f_{1.2}'(z)$ in form
Eq.(\ref{a.9}).

Entering in Eq.(\ref{a.9}) expression $(1+u)^{-1}$ we present as
contour integral
\begin{equation}
\frac{1}{(1+u)}=\frac{i}{2}\int_{\lambda-i\infty}^{\lambda+i\infty}\frac{u^s}{\sin\pi
s}ds,\quad u=\chi z^{3/2},\quad -1<\lambda<0. \label{a.10}
\end{equation}
Substituting in the integral in Eq.(\ref{a.9}) the functions
$f_{1,2}'(z)$ in the form given by the same equation and integrating
over the variables $z$ and $\tau$, we get after change of variable
$s \rightarrow 2s$, displacement of integration contour and
reduction of similar terms the new representation of the function
$J(\chi)$, which is suitable for both analytical and numerical
calculation:
\begin{equation}
J(\chi)=\frac{i\pi}{2}\int_{\lambda-i\infty}^{\lambda+i\infty}
\frac{\chi^{2s}}{3^s}
\frac{\Gamma(1+3s)}{\Gamma(s)}R(s)\frac{ds}{\sin^2\pi s},\quad
-\frac{1}{3} < \lambda <0 \label{a.14}
\end{equation}
where
\begin{equation}
R(s)=15+43s+31s^2+28s^3+12s^4. \label{a.15}
\end{equation}
In the case $\chi \ll 1$, closing the integration contour on the
right, one can calculate the asymptotic series in powers of $\chi$
\begin{equation}
J(\chi)=15+516\chi^2\left(\ln
\frac{\chi}{\sqrt{3}}-C\right)+1893\chi^2+\ldots \simeq
15\left[1-34.4\chi^2\left(\ln\frac{1}{\chi}-2.542\right)\right]
\label{a.16}
\end{equation}
Closing the integration contour on the left one obtains the series
over the inverse powers of $\chi$
\begin{equation}
J(\chi)=\frac{58\pi \Gamma(1/3)}{81\cdot 3^{1/6}\chi^{2/3}}+
\frac{628\pi3^{1/6} \Gamma(2/3)}{243 \chi^{4/3}}-\frac{13}{\chi^2}
\left(\ln \chi-\frac{1}{2}\ln 3-C+\frac{57}{52}\right)+\dots.
\label{a.17}
\end{equation}

Now we get over to the third term ($\propto \nu_0^4$) of the
decomposition. In this case it is convenient to turn back the
integration contour in Eq.(\ref{17}) and perform inverse
transformation $t \rightarrow t/\sqrt{2}$, so that $\sqrt{2}\nu_0 t
\rightarrow  \nu t~(\nu = \sqrt{i}\nu_0)$. In the terms $\propto
\nu_0^4$ of the decomposition the integrals over $t$ have the form
\begin{equation}
\int_0^{\infty}\exp\left(-i\left(t+\frac{a
t^3}{3}\right)\right)t^{2n+1}dt,\quad {\rm or} \quad
i\int_0^{\infty}\exp\left(-i\left(t+\frac{a
t^3}{3}\right)\right)t^{2n}dt,\label{d.1}
\end{equation}
where $a=\chi^2/u^2$. The radiation intensity is contains the
imaginary part of these integrals where the integrand is even
function of $t$. Because of this the final result is expressed in
terms of MacDonald functions $K_{1/3}(z)$ and
$K_{2/3}(z)~(z=2u/3\chi)$ and their derivatives. Using the
recurrence relations we find after quite cumbersome calculation
\begin{equation}
I^{(3)}= - \frac{\alpha m^2}{8400}
\frac{\varepsilon^2}{\varepsilon_0^2}g_r^2
\int_0^{x_0}e^{-2x/\eta_1}P(\chi)\frac{dx}{x_0},
\label{d.2}
\end{equation}
where
\begin{eqnarray}
\hspace{-10mm}&& P(\chi)=\frac{9\sqrt{3}}{64\pi}\int_0^{1}\left[r_2
F_2(z)+r_3 F_3(z)\right]z^3\frac{1-y}{y}dy,\quad u=\frac{y}{1-y},
\nonumber \\
\hspace{-10mm}&&
F_2(z)=(7820+126z^2)zK_{2/3}(z)-(280+2430z^2)K_{1/3}(z),\quad
F_3(z)=(264-63z^2)zK_{2/3}(z)-(24+3z^2)K_{1/3}(z).\label{d.3}
\end{eqnarray}

Passing on to the variable $u$ and having applied the representation
\begin{equation}
\frac{1}{(1+u)^m}=\frac{1}{2\pi i}
\int_{\lambda-i\infty}^{\lambda+i\infty}
\frac{\Gamma(-s)\Gamma(m+s)}{\Gamma(m)}u^s ds,\quad -m<\lambda<0,
\label{d.4}
\end{equation}
taking into account the table integrals over $u$
\begin{equation}
\int_0^{\infty}
x^{\mu}K_{\nu}(x)dx=2^{\mu-1}\Gamma\left(\frac{1+\mu+\nu}{2}
\right)\Gamma\left(\frac{1+\mu-\nu}{2}\right), \label{d.5}
\end{equation}
substituting $s \rightarrow 2s$ and using the tripling formula
\begin{equation}
3^{3s}\Gamma(s)\Gamma(s+1/3)\Gamma(s+2/3)=2\pi\sqrt{3}\Gamma(3s),
\label{d.6}
\end{equation}
we get after reduction of similar terms the following expression for
the function $P(\chi)$
\begin{eqnarray}
&&P(\chi)=\frac{1}{2\pi i} \int_{\lambda-i\infty}^{\lambda+i\infty}
D(s)
\left(\frac{\chi^2}{3}\right)^{s-1}(1-2s)\Gamma(1-s)\Gamma(3s)\frac{ds}{\cos(\pi
s)}, \quad 0<\lambda<1,
\nonumber \\
&& D(s)=192+532s-210s^2+73s^3-349s^4+42s^5. \label{d.7}
\end{eqnarray}
Closing the integration contour to the right we get the asymptotic
series over powers of $\chi$
\begin{equation}
P(\chi)=560+5985\sqrt{3}\chi - 388800\chi^2 +\ldots\label{d.8}
\end{equation}
Closing the integration contour to the left we get the series over
powers of $1/\chi$
\begin{equation}
P(\chi)=\frac{192}{\chi^2}
+\frac{4280}{243}3^{1/3}\Gamma\left(\frac{1}{3}\right)\frac{1}{\chi^{8/3}}-
\frac{635\sqrt{3}}{\chi^3}-
 +\ldots
 \label{d.9}
\end{equation}

The inverse radiation length in tungsten crystal (axis $<111>$)
$1/L^{cr}(\varepsilon)=I(\epsilon)/\varepsilon$ Eq.(\ref{17}), well
as coherent contribution
$1/L^F(\varepsilon)=I^F(\varepsilon)/\varepsilon$ Eq.(\ref{19}) and
incoherent contribution
$1/L^{inc}(\varepsilon)=I^{inc}(\varepsilon)/\varepsilon$
Eq.(\ref{21})  are shown in Fig.1 for two temperatures T=100 K and
T=293 K as a function of incident electron energy $\varepsilon$. In
low energy region ($\varepsilon \leq 0.3$~GeV) the asymptotic
expressions Eqs.(\ref{7}) and (\ref{8}) are valid. One can see that
at temperature T=293 K the intensity $I^F(\varepsilon)$ is equal to
$I^{inc}(\varepsilon)$ at $\varepsilon \simeq 0.4$~GeV and
temperature T=100 K the intensity $I^F(\varepsilon)$ is equal to
$I^{inc}(\varepsilon)$ at $\varepsilon \simeq 0.7$~GeV. The same
estimates follow from comparison of Eqs.(\ref{7}) and (\ref{8}), see
also Eq.(\ref{10}). At higher energies the intensity
$I^F(\varepsilon)$ dominates while the intensity
$I^{inc}(\varepsilon)$ decreases monotonically.

The inverse radiation length given in Fig.1 can be compared with
data directly only if the crystal thickness $l \ll
L^{cr}(\varepsilon)$ (thin target). Otherwise one has to take into
account the energy loss. The corresponding analysis is simplified
essentially if $l \leq L^{min}=({\rm max}
(I(\varepsilon)/\varepsilon))^{-1}$. The radiation length
$L^{cr}(\varepsilon)$ varies slowly on the electron trajectory for
such thicknesses. This is because of weak dependence of
$L^{cr}(\varepsilon)$ on energy in the region $L^{cr}(\varepsilon)
\simeq L^{min}$ and the relatively large value of
$L^{cr}(\varepsilon) \gg L^{min}$ in the region where this
dependence is essential but variation of energy on the thickness $l$
is small. For W, axis $<111>$, T=293 K one has $L^{min}=320~\mu m$
at energy $\varepsilon=300$~GeV, see Fig.1. For this situation
dispersion can be neglected (see discussion in Sec.17.5 of
\cite{BKS}) and energy loss equation acquires the form
\begin{equation}
\frac{1}{\varepsilon}\frac{d\varepsilon}{dl}=
-L^{cr}(\varepsilon)^{-1}\equiv -\frac{I(\varepsilon)}{\varepsilon}.
\label{18a}
\end{equation}

\begin{figure}[h]
\begin{picture}(170,100)
\put(42,50){\makebox(0,0){\includegraphics[width=76mm]{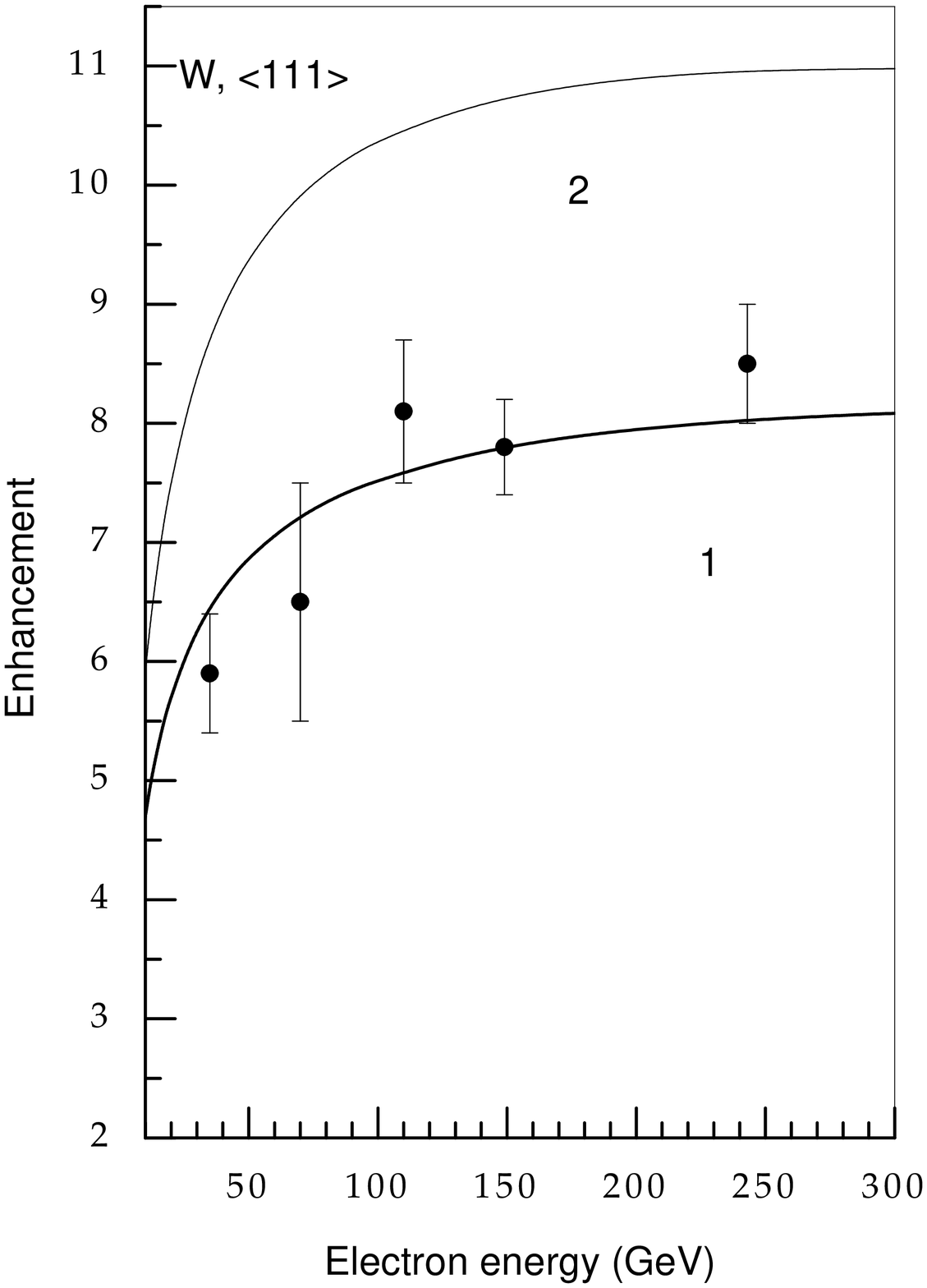}}}
\put(127,50){\makebox(0,0){\includegraphics[width=70mm]{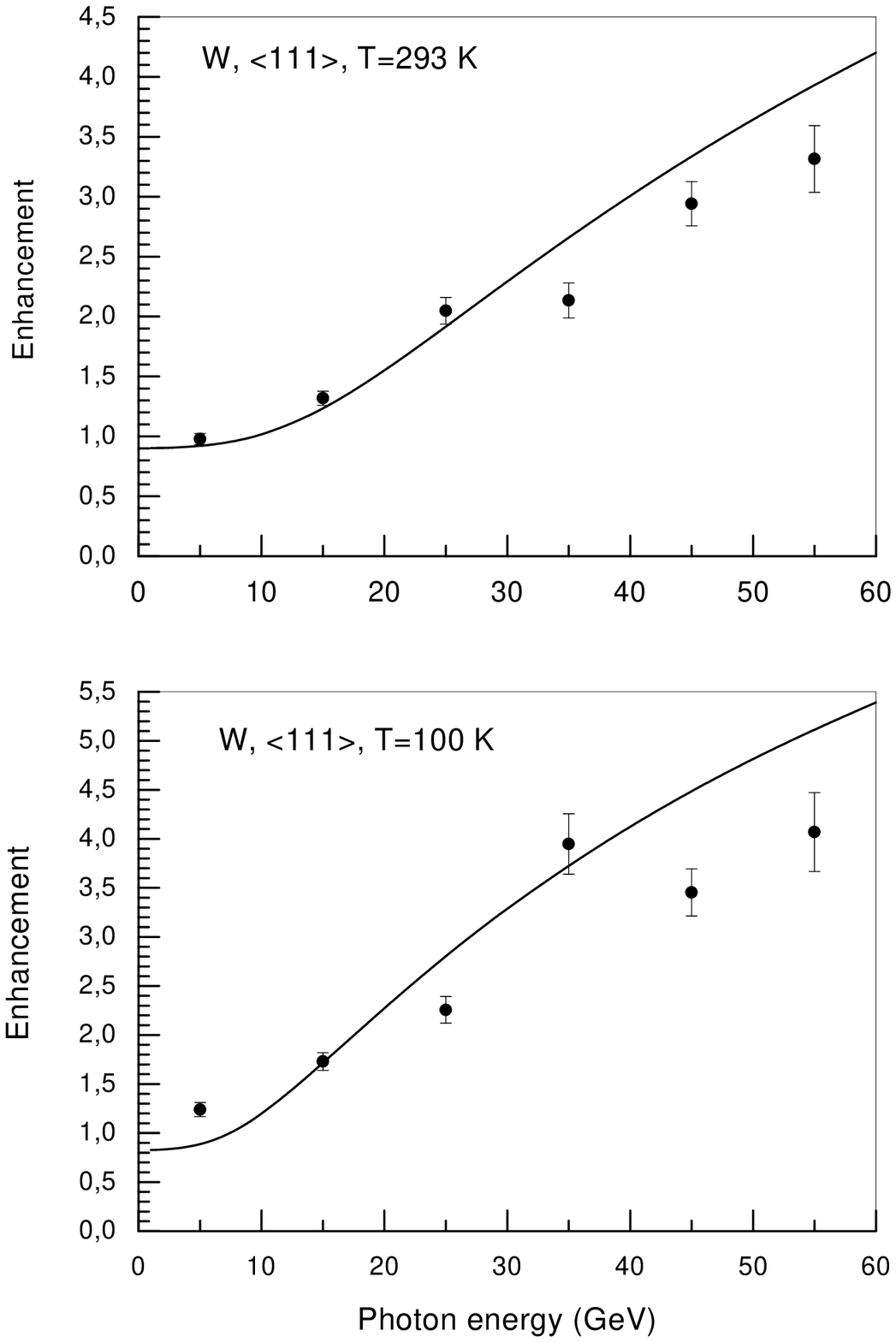}}}
\put(42,108){\makebox(0,0){(a)}} \put(127,108){\makebox(0,0){(b)}}
\end{picture}
\caption{Comparison of theory and experiment.(a)~Enhancement of
radiation intensity (the ratio $L^{BM}/L^{ef}$) in tungsten, axis
$<111>$, T=293 K. The curve 1 is for the target with thickness
$l=200~\mu m$, where the energy loss was taken into account
(according with Eq.(\ref{18e})). The curve 2 is for a considerably
more thinner target, where one can neglect the energy loss ($L^{ef}
\rightarrow L^{cr}$). The data are from \cite{KMU}.
\newline (b)~Enhancement of the probability of pair creation in tungsten
for different temperatures, axis $<111>$. The data are from
\cite{KKM}.}
\end{figure}

In the first approximation the final energy of electron is
\begin{equation}
\varepsilon_1=\varepsilon_0
\exp\left(-l/L^{cr}(\varepsilon_0)\right), \label{18b}
\end{equation}
where $\varepsilon_0$ is the initial energy. In the next
approximation one has
\begin{equation}
\ln\frac{\varepsilon(l)}{\varepsilon_0}=-L^{cr}(\varepsilon_0)
\int_{\varepsilon_1}^{\varepsilon_0}L^{cr}(\varepsilon)^{-1}
\frac{d\varepsilon}{\varepsilon}. \label{18c}
\end{equation}
If the dependence of $L^{cr}(\varepsilon)^{-1}$ on $\varepsilon$ is
enough smooth it's possible to substitute the function
$L^{cr}(\varepsilon)^{-1}$ by an average value with the weight
$1/\varepsilon$:
\begin{equation}
L^{cr}(\varepsilon)^{-1} \rightarrow
\frac{\varepsilon_0L^{cr}(\varepsilon_1)^{-1}+\varepsilon_1L^{cr}(\varepsilon_0)^{-1}}
{\varepsilon_0+\varepsilon_1}\equiv \frac{1}{\overline{L}}.
\label{18d}
\end{equation}
Numerical test confirms this simplified procedure. Using it we find
\begin{equation}
\ln\frac{\varepsilon(l)}{\varepsilon_0}=
-\frac{L^{cr}(\varepsilon_0)}{\overline{L}}
\ln\frac{\varepsilon_0}{\varepsilon_1}=-\frac{l}{\overline{L}},\quad
\frac{\Delta
\varepsilon}{\varepsilon_0}=1-\exp\left(-\frac{l}{\overline{L}}\right)
\equiv \frac{l}{L^{ef}}. \label{18e}
\end{equation}

Enhancement of  radiation length (the ratio of Bethe-Maximon
radiation length $L^{BM}$ and $L^{ef}$ ) in tungsten, axis $<111>$,
T=293 K is shown in Fig.2(a). The curve 1 is for the target with
thickness $l=200~\mu m$, where the energy loss was taken into
account according using the simplified procedure Eq.(\ref{18e}). The
curve 2 is for a considerably more thinner target, where one can
neglect the energy loss. The only available data are from
\cite{KMU}. The measurement of radiation from more thin targets is
of evident interest.

\textbf{\fontsize{10}{12pt}\selectfont 3.2.~Pair creation}

The general expression for integral probability of pair creation by
a photon under the simultaneous action of multiple scattering and an
external constant field was obtained in \cite{BK} (see Eqs.(2.14)
and (1.12)). This expression can be found also from Eq.(\ref{17})
using the standard QED substitution rules: $\varepsilon \rightarrow
-\varepsilon,~\omega \rightarrow -\omega,~\omega^2d\omega
\rightarrow \varepsilon^2d\varepsilon$ and exchange $
\varepsilon_c(x) \rightarrow \omega_c(x)/4$;
\begin{eqnarray}
&& W=\frac{\alpha m^2}{2\pi \omega}\int_0^1 \frac{dy}{y(1-y)}
\int_0^{x_0}\frac{dx}{x_0}G(x, y),\quad G(x, y)=\int_0^{\infty} F(x,
y, t)dt +s_3\frac{\pi}{4},
\nonumber \\
&& F(x, y, t)={\rm Im}\left\lbrace e^{f_1(t)}\left[s_2\nu_0^2
(1+ib)f_2(t)-s_3f_3(t) \right] \right\rbrace,\quad
b=\frac{4\kappa_1^2}{\nu_0^2}, \quad y=\frac{\varepsilon}{\omega},
\nonumber \\
&& f_1(t)=(i-1)t+b(1+i)(f_2(t)-t),\quad
f_2(t)=\frac{\sqrt{2}}{\nu_0}\tanh\frac{\nu_0t}{\sqrt{2}},\quad
f_3(t)=\frac{\sqrt{2}\nu_0}{\sinh(\sqrt{2}\nu_0t)}, \label{15p}
\end{eqnarray}
where
\begin{equation}
s_2=y^2+(1-y)^2,~s_3=2y(1-y),~\nu_0^2=4y(1-y)
\frac{\omega}{\omega_c(x)},~\kappa_1=y(1-y)\kappa(x), \label{16p}
\end{equation}
$\varepsilon$ is the energy of one of the particles of pair, the
function $\omega_c(x)$ is defined in Eq.(\ref{11p}) and $\kappa(x)$
is defined in Eq.(\ref{5}).

\begin{figure}[h]
\begin{picture}(170,68)
\put(32,32){\makebox(0,0){\includegraphics[width=64mm]{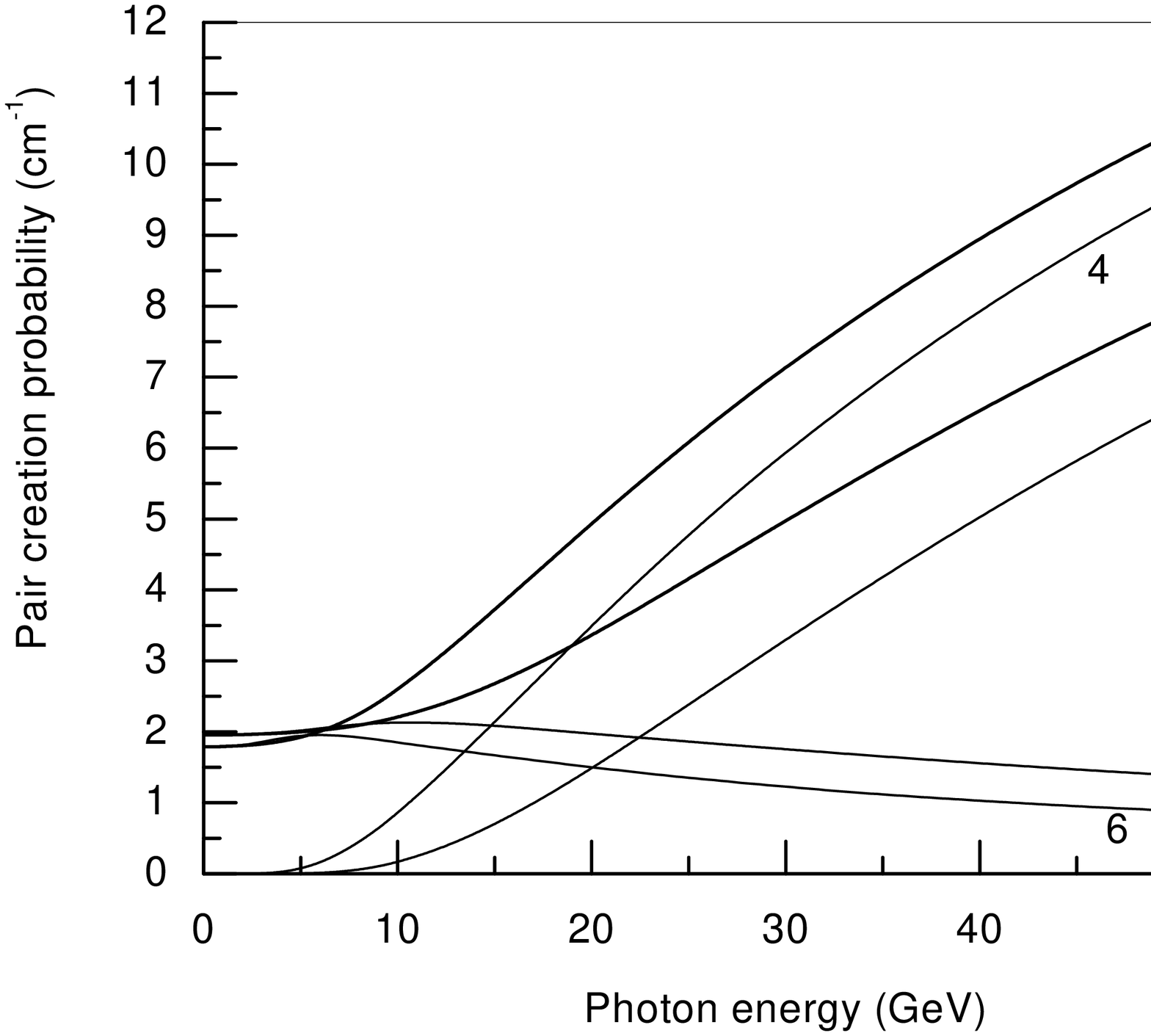}}}
\put(127,32){\makebox(0,0){\includegraphics[width=82mm]{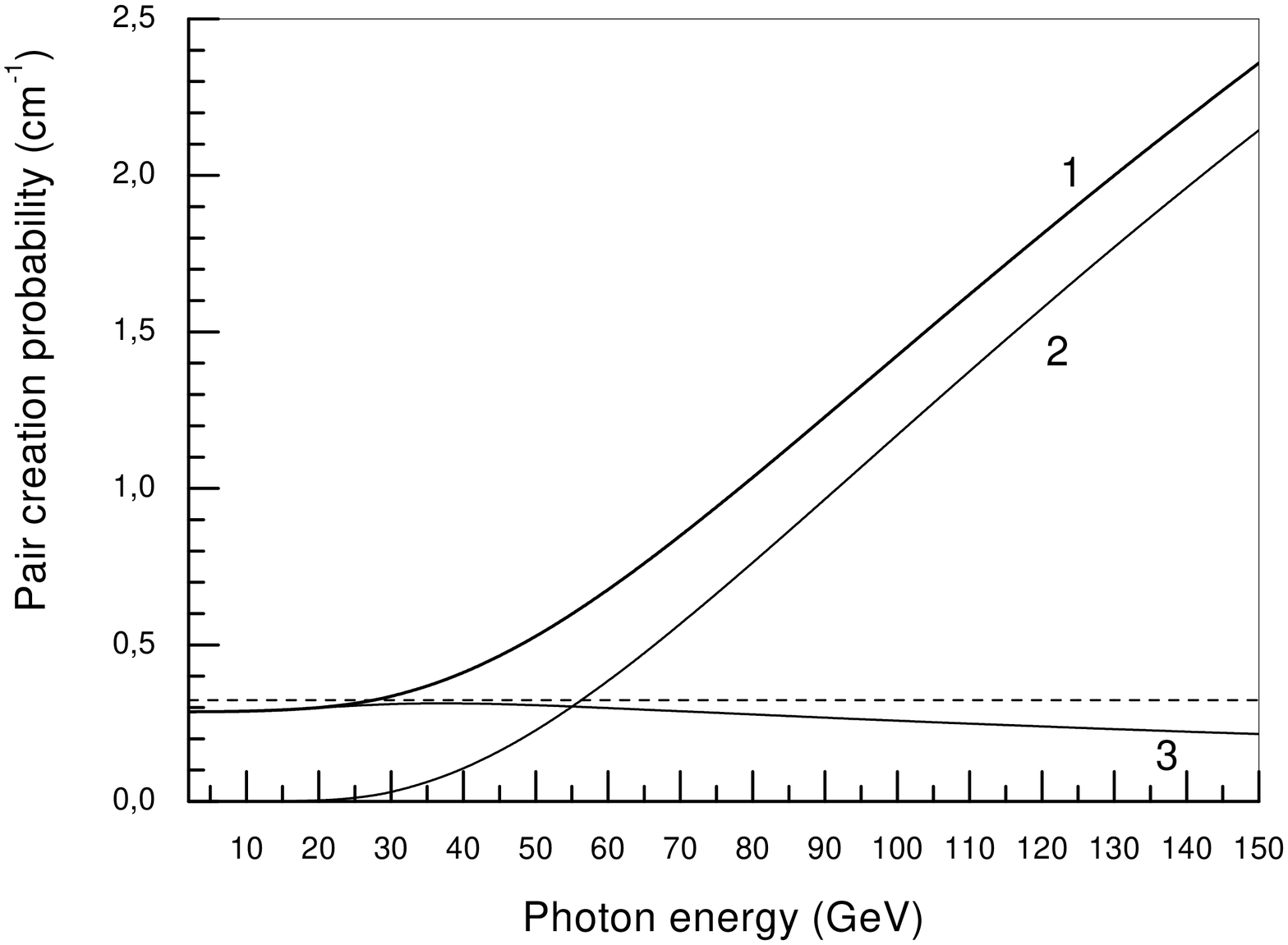}}}
\put(32,64){\makebox(0,0){(a)}} \put(127,64){\makebox(0,0){(b)}}
\end{picture}
\caption{(a)~Pair creation probability in tungsten, axis $<111>$ at
different temperatures T. Curves 1 and 3 are the total probability
$W$ Eq.(\ref{15p}) for T=293 K and T=100 K, the curves 2 and 4 give
the coherent contribution $W^F$ Eq.(\ref{17p}), the curves 5 and 6
give the incoherent contribution $W^{inc}$ Eq.(\ref{18p}) at
corresponding temperatures T. \newline (b)~Pair creation probability
in germanium , axis $<110>$ at T=100 K. Curve 1 is the total
probability $W$ Eq.(\ref{15p}), the curve 2 gives the coherent
contribution $W^F$ Eq.(\ref{17p}), the curve 3 gives the incoherent
contribution $W^{inc}$ Eq.(\ref{18p}). The dashed line is the
Bethe-Maximon probability}
\end{figure}

In order to single out the influence of the multiple scattering (the
LPM effect) on the process under consideration, we should consider
both the coherent and incoherent contributions. The probability of
coherent pair creation is the first term ($\nu_0^2=0$) of the
decomposition of Eq.(\ref{15p}) over $\nu_0^2$ (compare with
Eq.(2.17) in \cite{BK} and see Eq.(12.7) in \cite{BKS})
\begin{eqnarray}
&&W^F=\frac{\alpha m^2}{2\sqrt{3}\pi\omega}\int_0^1\frac{dy}{y(1-y)}
\int_0^{x_0}\frac{dx}{x_0}\Bigg[2s_2K_{2/3}(\lambda)\nonumber \\
&&+s_3\int_{\lambda}^{\infty}K_{1/3}(z)dz \Bigg],\quad
\lambda=\frac{2}{3\kappa_1}. \label{17p}
\end{eqnarray}

The probability of incoherent pair creation is the second term
($\propto \nu_0^2$) of the mentioned decomposition (compare with
Eq.(2.26) in \cite{BK} and compare with Eq.(21.31) in \cite{BKS})
\begin{equation}
W^{inc}=\frac{4Z^2\alpha^3 n_a L}{15 m^2}\int_0^1 dy
\int_0^{\infty}\frac{dx}{\eta_1}e^{-x/\eta_1}f(x, y), \label{18p}
\end{equation}
where $L$ is defined in Eq.(\ref{10}),
\begin{equation}
f(x, y)=f_1(z)+ s_2 f_2(z),\quad z=z(x, y)=\kappa_1^{-2/3},
\label{19p}
\end{equation}
here functions $f_{1,2}(z)$ are defined in Eq.(\ref{a.11a}). In
further analysis and numerical calculation it is convenient to use
given above representations of the Hardy function and its derivative
Eqs.(\ref{a.12a}),(\ref{a.6}).

The probabilities $W$ Eq.(\ref{15p}), $W^F$ Eq.(\ref{17p}), and
$W^{inc}$ Eq.(\ref{18p}) at different temperatures T are shown in
Fig.3 as a function of photon energy $\omega$. In low energy region
($\omega \leq 1$~GeV) one can neglect the coherent process
probability $W^F$ as well as influence of axis field on the
incoherent process probability and the LPM effect and the
probability of process is $W^{LE}=n_a\sigma_p$ Eq.(\ref{7p}). In
this energy region as one can see in Fig.3  the probability $W$ is
by 10\% at T=293 K and  by 20\% at T=100 K less than the probability
at random orientation $W^{ran}$ which is taken as $W^{ran}=W^{BM}$
(the Bethe-Maximon probability is $W^{BM}=W_0(1-1/42L_0)$=2.17 1/cm
in tungsten).

With energy increase the influence of axis field begins and the LPM
effect manifests itself according to Eq.(\ref{13p}) (the terms with
$\overline{\kappa^2}$ and $(\omega g/\omega_0)^2$ correspondingly).
This leads first to not large increase of the probability $W^{inc}$
which attains the maximum at $\omega \sim \omega_m$. The probability
$W^F$ in this region is defined by Eq.(\ref{5a}) and its
contribution is relatively small. The probability $W^F$ becomes
comparable with $W^{inc}$ at $\omega \simeq 1.5\omega_m$. At higher
energies $W^F$ dominates, while $W^{inc}$ decreases monotonically.

In Fig.2(b) the calculated total integral probability $W$ of pair
creation by a photon Eq.(\ref{15p}) is compared with data of NA43
CERN experiment \cite{KKM}. The enhancement is the ratio $W/W^{BM}$.
One can see that the theory quite satisfactory describes data. This
statement differs from conclusion made in \cite{KKM}. One of reasons
for this difference is diminishing of incoherent contribution (see
Fig.3): for W, $<111>$, T=100 K at photon energy $\omega=55$~GeV one
has $W^{inc}=0.35W^{BM}$, while in \cite{KKM} it was assumed that
$W^{inc}=W^{BM}$.

The third term ($\propto \nu_0^4$) of the decomposition the pair
creation probability  over $\nu_0^2$ can be obtained from the
corresponding expressions for radiation using the QED standard
substitutions (cp with Eq.(\ref{15p})) and taking into account that
$\omega_0=4\varepsilon_0$ (see Table 1)
\begin{eqnarray}
&& W^{(3)}= - \frac{\alpha m^2}{8400} \frac{\omega}{\omega_0^2}g^2
\int_0^{x_0}e^{-2x/\eta_1}T(\kappa)\frac{dx}{x_0},
\nonumber \\
&& T(\kappa)=\frac{9\sqrt{3}}{4\pi}\int_0^{1}\left[s_2
F_2(\lambda)-s_3 F_3(\lambda)\right]\lambda^3y(1-y)dy \label{d.7p}
\end{eqnarray}

\begin{center}
\textbf{\fontsize{11}{13pt}\selectfont 3.~ THE LPM EFFECT IN
ORIENTED CRYSTAL}
\end{center}
\vspace{2mm}

The contribution of the LPM effect in the total intensity of
radiation $I$ Eq.(\ref{17}) is defined as
\begin{equation}
I^{LPM}=I - I^F -I^{inc} \label{23}
\end{equation}
The relative contribution (negative since the LPM effect suppresses
the radiation process) $\Delta_r=-I^{LPM}/I$ is shown in Fig.4(a).
This contribution has the maximum $\Delta_r \simeq 0.8$\% at
$\varepsilon \simeq 0.7$~GeV for T=293 K and $\Delta_r \simeq 0.9$\%
at $\varepsilon \simeq 0.3$~GeV for T=100 K or, in general, at
$\varepsilon \sim \varepsilon_t$. The left part of the curves is
described quite satisfactory by  Eq.(\ref{16}).  For explanation of
the right part of the curves let us remind that at $\varepsilon \gg
\varepsilon_m$ the behavior of the radiation intensity at $x \sim
\eta_1$ is defined by the ratio of the contributions to the momentum
transfer of multiple scattering and that of the external field on
the formation length $l_f$ (see Eq.(21.3) in \cite{BKS})
\begin{eqnarray}
&&k=\frac{<{\bf q}_s^2>}{<{\bf q}>^2}
=\frac{\dot{\vartheta}_s^2l_f}{(wl_f)^2}\sim
\frac{\varepsilon}{\varepsilon_0}
\chi_m^{-4/3}=\frac{\varepsilon}{\varepsilon_0}
\left(\frac{\varepsilon_m}{\varepsilon}\right)^{4/3},
\nonumber \\
&&\frac{1}{L^F} \sim \frac{\alpha}{l_f} \sim \frac{\alpha
m^2}{\varepsilon}\chi_m^{2/3} = \frac{\alpha
m^2}{\varepsilon_m}\chi_m^{-1/3}, \label{24}
\end{eqnarray}

\begin{figure}[h]
\begin{picture}(170,68)
\put(38,32){\makebox(0,0){\rotatebox{270}{\includegraphics[width=59.5mm]{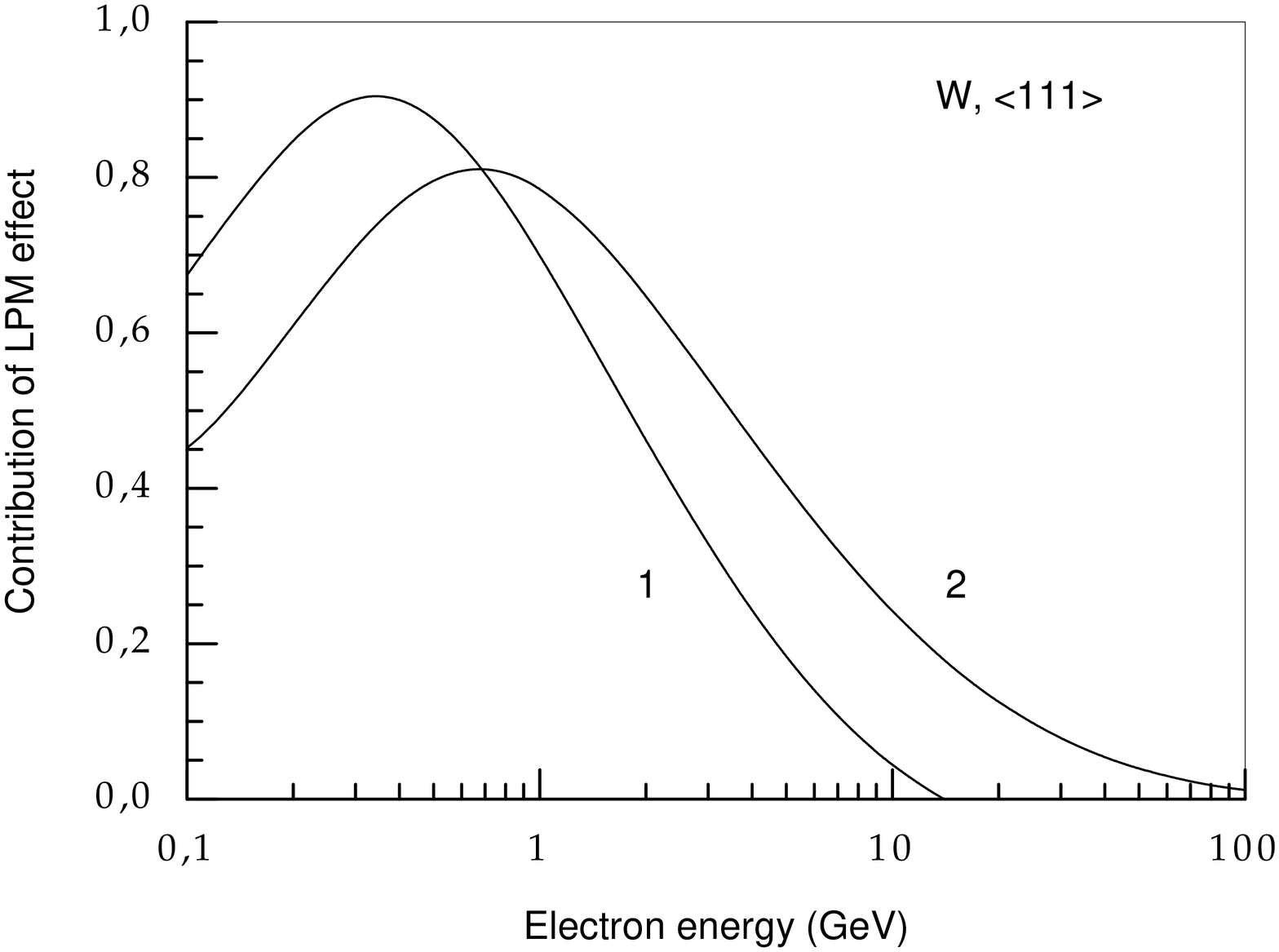}}}}
\put(127,32){\makebox(0,0){\includegraphics[width=80mm]{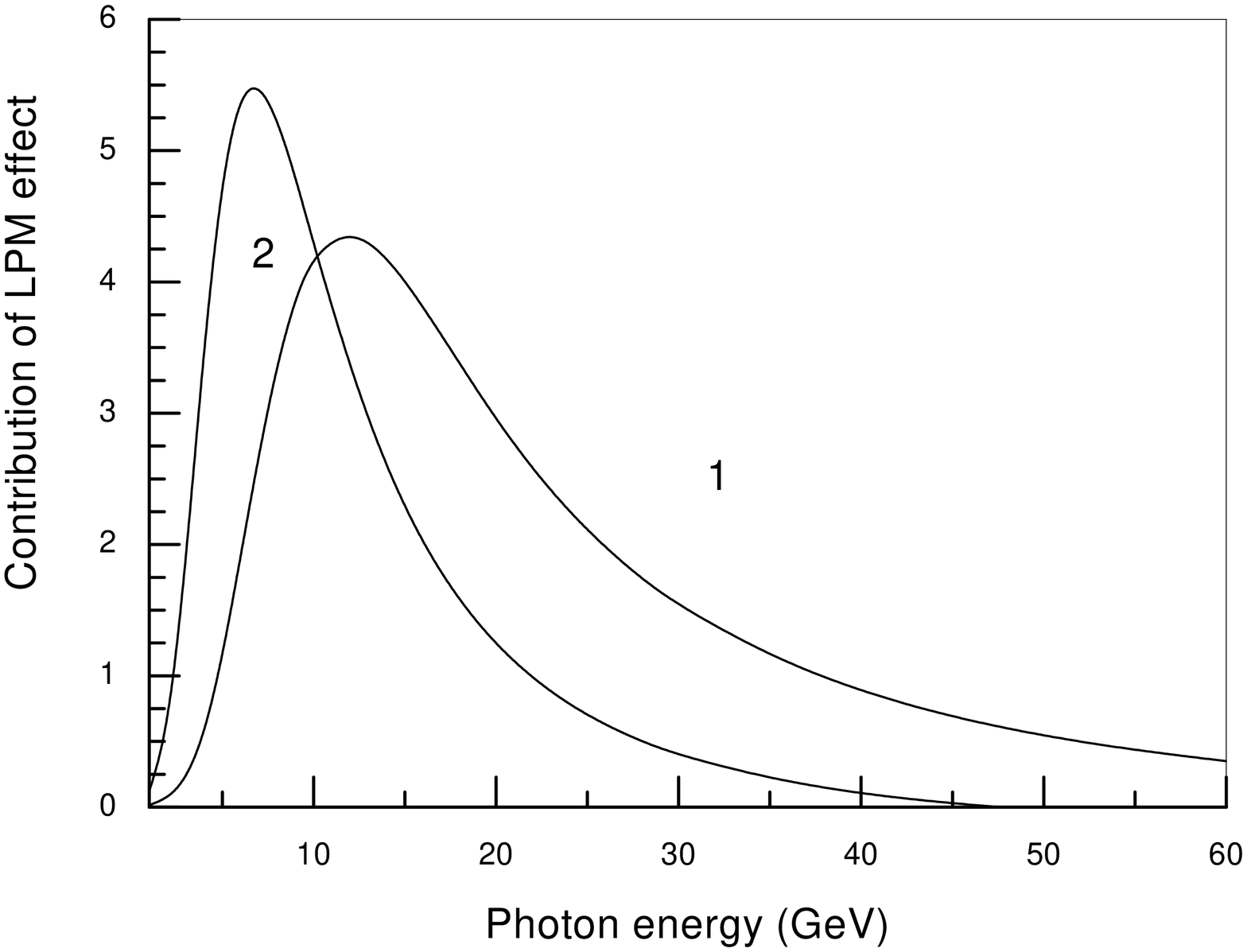}}}
\put(38,65){\makebox(0,0){(a)}} \put(127,64){\makebox(0,0){(b)}}
\end{picture}
\caption{The LPM effect in oriented crystal:(a)~The relative
contribution into radiation intensity of the LPM effect $\Delta_r$
(per cent) in tungsten, axis $<111>$. Curve 1 is for T=100 K and
curve 2 is for T=293 K.\newline (b)~The relative contribution of the
LPM effect into total pair creation probability $\Delta$ (per cent)
in tungsten, axis $<111>$. Curve 1 is for T=293 K and curve 2 is for
T=100 K. }
\end{figure}
where $w$ is an acceleration in an external field. The linear over
$k$ term determines the contribution into intensity of incoherent
process: $1/L^{inc}(\varepsilon \gg \varepsilon_m) \sim
k/L^F(\varepsilon) \sim \alpha m^2/(\varepsilon_0 \chi_m^{2/3})$.
The LPM effect is defined by the next term of decomposition over
$k~(\propto k^2)$ and decreases with energy even faster than
$1/L^{inc}(\varepsilon)$. Moreover one has to take into account that
at $\varepsilon \geq \varepsilon_s$ the contribution of relevant
region $x \sim \eta_1$ into the total radiation intensity is small
and $1/L^F(\varepsilon)$ decreases with the energy growth as
$\chi_m^{-1/3}$. For such energies the main contribution gives the
region $x \sim \chi_s^{2/3}=(\varepsilon/\varepsilon_s)^{2/3}$ and
$1/L^{cr}(\varepsilon)$ increases until energy $\varepsilon \sim 10
\varepsilon_s$ (see Fig.1). This results in essential reduction of
relative contribution of the LPM effect $\Delta_r$.

\begin{wrapfigure}{l}{0.42\textwidth}
\includegraphics[width=0.41\textwidth]{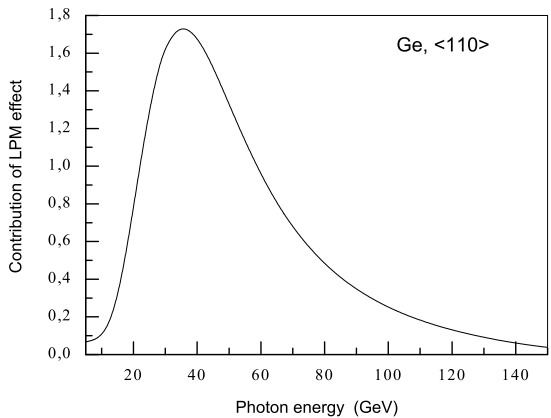}
\caption{ The relative contribution of the LPM effect into total
pair creation probability $\Delta$ (per cent) in germanium crystal,
axis $<110>$, for T=100 K. }
\end{wrapfigure}

The contribution of the LPM effect in the total pair creation
probability $W$ Eq.(\ref{15p}) is defined as
\begin{equation}
W^{LPM}=W - W^F -W^{inc} \label{23p}
\end{equation}
The relative contribution (negative since the LPM effect suppresses
the process) $\Delta=-W^{LPM}/W$ is shown in Fig.4(b). This
contribution has the maximum $\Delta \simeq 5.5$\% at $\omega \simeq
7$~GeV for T=293 K and $\Delta \simeq 4.3$\% at $\omega \simeq
12$~GeV for T=100 K for tungsten crystal. For germanium crystal the
value $\Delta=-W^{LPM}/W$ is shown in Fig.5. This contribution
attains the maximum  $\Delta \simeq 1.7$\% at $\omega \simeq 34$~GeV
for T=100 K or, in general, at $\omega \sim\omega_m$. The left part
of the curves is described by the term with $(\omega g/\omega_0)^2$
in Eq.(\ref{13p}). For understanding of the right part of the curves
one has to take into account that at $\omega \gg \omega_m$ the
behavior of the pair creation probability at $x \sim \eta_1$ is
defined by the ratio of the contributions to the momentum transfer
of multiple scattering and that of the external field on the
formation length $l_f$(see Eqs.(1.4), (2.28), (2.29) and discussion
in \cite{BK})
\begin{eqnarray}
&&k=\frac{<{\bf q}_s^2>}{<{\bf q}>^2}
=\frac{\dot{\vartheta}_s^2l_f}{(wl_f)^2}\sim \frac{\omega}{\omega_0}
\kappa_m^{-4/3}=\frac{\omega}{\omega_0}
\left(\frac{\omega_m}{\omega}\right)^{4/3},
\nonumber \\
&& W^F \sim \frac{\alpha}{l_f} \sim \frac{\alpha
m^2}{\omega}\kappa_m^{2/3}, \label{24p}
\end{eqnarray}
where $w$ is an acceleration in an external field. The linear over
$k$ term determines the contribution into probability of incoherent
process: $W^{inc}(\omega \gg \omega_m) \sim k W^F \sim \alpha
m^2/(\omega_0 \kappa_m^{2/3})$ (cp Eq.(\ref{9p})). The LPM effect is
defined by the next term of decomposition over $k~(\propto k^2)$ and
decreases with energy even faster than $W^{inc}$.

It follows from Eq.(\ref{24p}) that maximal influence of multiple
scattering on the process under consideration is reached at $\omega
\sim \omega_m \sim \omega_0$ where $k \sim 1$ for tungsten crystal.
The analysis shows (see Fig.4 in \cite{BK5}) that at $\omega \sim
\omega_e$ the LPM effect results in 10\% suppression of total
(incoherent) pair creation probability. In oriented crystal
$\omega_e \rightarrow \omega_0$ and at $\omega \sim \omega_m$ the
coherent and incoherent contributions are nearly equal. So, at this
photon energy one can expect $\sim 5$\% LPM effect for pair creation
process. This perfectly agrees with performed numerical calculation.
For germanium crystal $k \sim \omega_m/\omega_0 \sim 1/5$ at $\omega
\sim \omega_m$ (see discussion after Eq.(\ref{5a})) and for this
energy one can expect the LPM effect of order $\sim 1$\% for pair
creation process (the term with $(\omega g/\omega_0)^2$ in
Eq.(\ref{13p})).

In just the same way the maximal influence of multiple scattering on
the  incoherent radiation process (see Eq.(\ref{24})) is reached in
heavy elements, e.g. tungsten, at $\varepsilon \sim \varepsilon_m
\sim \varepsilon_0$ where $k \sim 1$. However at $\varepsilon \sim
\varepsilon_m $ the intensity of incoherent radiation constitutes
only one tenth of coherent contribution. Owing to this the maximum
of the LPM effect manifestation in the radiation process is shifted
to the left up to $\varepsilon \sim \varepsilon_t $, where the
coherent and incoherent contributions to the radiation intensity are
nearly equal and $\nu_0^2 \sim \varepsilon_t/\varepsilon_0 \sim
1/10$. This explains essentially smaller influence of the LPM effect
on radiation process.

\begin{center}
\textbf{\fontsize{11}{13pt}\selectfont 4.~ CONCLUSIONS}
\end{center}
\vspace{2mm}

So the rather prevalent assumption that the LPM effect can
essentially suppress the radiation and pair creation process in
oriented crystals is proved wrong due to action of axis field. On
the other hand, the LPM effect can be observed in accurate
measurements. For observation the LPM effect of mentioned scale in
an amorphous tungsten in hard part of the spectrum of radiation
process the electrons with the energy $\varepsilon \simeq 2.5$~TeV
are needed (or for pair creation process the photons with energy
$\omega \simeq 10$~TeV are needed) \cite{BK1}.

So in high energy region the mechanisms of radiation and pair
creation by a photon are very different in an amorphous medium and
in oriented crystal. In amorphous medium the radiation intensity is
suppressed substantially at $\varepsilon > \varepsilon_e$ (or the
probability of pair creation is suppressed substantially at $\omega
> \omega_e$) due to the LPM effect and tends to zero at
$\varepsilon \gg \varepsilon_e$ (or $\omega \gg \omega_e$)
\cite{BK1}. In oriented crystal the coherent mechanism dominates and
at $\chi_s \gg 1$ (or $\kappa_s \gg 1$) the radiation intensity (or
probability of pair creation) is decreasing also (see Eq.(17.17) (or
Eq.(12.16)) in \cite{BKS}). The incoherent mechanism is suppressed
and the LPM effect is suppressed more strongly as it is follow from
the above discussion. It should be noted that the radiation
intensity (and the probability of pair creation) in oriented crystal
is always much higher than in a corresponding amorphous medium.

It's instructive to compare the LPM effect in oriented crystal for
radiation and pair creation processes. The manifestation of the LPM
effect is essentially different because of existence of threshold in
pair creation process. The threshold energy $\omega_m$ is relatively
high (in W, axis $<111>$, $\omega_m \sim 8$~GeV for T=100 K and
$\omega_m \sim 14$~GeV for T=293 K). Below $\omega_m$ influence of
field of axis is weak and the relative contribution of the LPM
effect attains 5.5 \% for T=100 K \cite{BK0}. There is no threshold
in radiation process and $I^F$ becomes larger than  $I^{inc}$ at
much lower energy $\varepsilon_t$ and starting from this energy the
influence of field of axis suppresses strongly the LPM effect. So
the energy interval in which the LPM effect could appear is much
narrower than for pair creation and its relative contribution is
less than 1 \% in W, axis $<111>$. Since value of $\varepsilon_t$
depends weakly on $Z$ (Eq.(\ref{10})), $\varepsilon_m \propto
Z^{-1}$ (Eqs.(\ref{5}), (\ref{6})) and $\varepsilon_0 \propto
Z^{-2}$ (Eq.(\ref{9})) the relative contribution of the LPM effect
$\Delta$ for light elements significantly smaller. Thus, the above
analysis shows that influence of multiple scattering on basic
electromagnetic processes in oriented crystal (radiation and pair
creation) is very limited especially for radiation process.

Let us note the important result obtained connected with
decomposition of Eqs.(\ref{17}) and (\ref{15p}) over powers of
$\nu_0^2$. The above analysis shows that the characteristics of the
processes under consideration are described quite satisfactory by
the two first terms of the decomposition over $\nu_0^2$ (the
coherent and incoherent contributions). The applicability of the
third term of the decomposition ($\propto \nu_0^4$) is restricted to
either very low energy interval (see Eqs.(\ref{8}) and (\ref{8p})
and corresponding comments) or very high energy region (because a
weak dependence of $k$ on $\varepsilon$ ($\omega$) since $k \propto
\varepsilon^{1/3}$). In the both limiting cases the LPM effect is
negligibly small, but in the energy interval where the LPM effect
could manifest itself, one has to apply the general formulas
Eqs.(\ref{17}) and (\ref{15p}).

{\bf Acknowledgments}

We are grateful to U.Uggerhoj for data. The authors are indebted to
the Russian Foundation for Basic Research supported in part this
research by Grant 06-02-16226.

\end{document}